  \documentclass[12pt]{article}
  \usepackage{amssymb}
  \usepackage{latexsym}

 \catcode`\@=11 \@addtoreset{equation}{section}\catcode`\@=12

\newcommand{\R}{ {\mathbb R} }

\newcommand{\fnm}{\footnotemark}
\newcommand{\fnt}{\footnotetext}

\begin{document}

 \begin{center}

 \large \bf

 On multidimensional solutions in the Einstein-Gauss-Bonnet model with a cosmological term

           \end{center}

 \vspace{0.3truecm}

 \begin{center}

   \bf A.A. Kobtsev\fnm[1]\fnt[1]{e-mail: aak@inr.ru}, 
    V.D. Ivashchuk\fnm[2]\fnt[2]{e-mail:
    ivashchuk@mail.ru} and  K.K. Ernazarov\fnm[3]\fnt[3]{e-mail:
    kubantai80@mail.ru}

\vspace{0.3truecm}

 \it  Moscow Mesons Factory of INR RAS,
      Moscow, Troitsk, 142190,  Russia; 
 
 \it   Center for Gravitation and Fundamental Metrology,
     VNIIMS, 46 Ozyornaya ul., Moscow, 119361, Russia;

 \it   Institute of Gravitation and Cosmology,
   Peoples' Friendship University of Russia,
   6 Miklukho-Maklaya ul., Moscow, 117198, Russia.

 \end{center}

 \begin{abstract}

  A $D$-dimensional  gravitational model with Gauss-Bonnet and cosmological term $\Lambda$ is
  considered. When  ansatz with diagonal  cosmological  metrics is adopted,
  we overview recent solutions for $\Lambda = 0$  and 
  find  new examples of solutions for $\Lambda \neq 0$ and $D=8$ with exponential dependence of scale factors
  which  describe   an  expansion of ``our'' 3-dimensional factor-space
  and  contraction of 4-dimensional internal space.

 \end{abstract}

  \newpage

%%%%%%%%%%%%%%%%%%%%%%%%%%%%%%%%%%%%%%%%%%%
\section{Introduction}
%%%%%%%%%%%%%%%%%%%%%%%%%%%%%%%%%%%%%%%%%%

Here we consider $D$-dimensional gravitational model with the Gauss-Bonnet term.
The action reads
\begin{equation}
\label{act}
 S =  \int_{M} d^{D}z \sqrt{|g|} \{ \alpha_1 (R[g] - 2 \Lambda) +
             \alpha_2  {\cal L}_2[g] \},
   \label{1.1}
 \end{equation}
where $g = g_{MN} dz^{M} \otimes dz^{N}$ is the metric defined on
the manifold $M$, ${\dim M} = D$, $|g| = |\det (g_{MN})|$ and

\begin{equation}
  {\cal L}_2 = R_{MNPQ} R^{MNPQ} - 4 R_{MN} R^{MN} +R^2
   \label{1.2}
 \end{equation}
is the standard Gauss-Bonnet term. Here $\alpha_1$ and $\alpha_2$
are non-zero constants.

 Earlier the appearance of the  Gauss-Bonnet
term  was motivated by string theory
 \cite{Zwiebach,GBstrings1,GBstrings2,GBstrings3,GBstrings4}.

 At present, the (so-called) Einstein-Gauss-Bonnet (EGB) gravitational model and
 its modifications are intensively used in  cosmology,
 see \cite{NojOd0} (for $D =4$),
 \cite{Ishihara,Deruelle,ElMakObOsFil,BambaGuoOhta,TT,KirMPTop,PTop,KirMak,ChPavTop}
 and references therein,  e.g. for explanation  of  accelerating
 expansion of the Universe following from
 supernovae (type Ia) observational data \cite{Riess,Perl,Kowalski}.
 Certain exact solutions in multidimesional EGB cosmology
 were obtained in  \cite{Ishihara}-\cite{ChPavTop}  and
 some other papers.

  Here we are dealing with the cosmological type solutions with diagonal
 metrics (of Bianchi-I-like type) governed by $n$ scale factors
 depending upon one variable, where $n > 3$. Moreover, we restrict ourselves by the
 solutions with exponential dependence of scale factors. We present new examples 
 of exact solutions in dimension $D=8$ which  describe   an exponential expansion 
 of 3-dimensional factor-space and contraction of 4-dimensional internal space.

%%%%%%%%%%%%%%%%%%%%%%%%%%%%%%%%%%%%%%%%%%%%%%%%%%%%%%%%%%%%%%%%%%
\section{The Cosmological  Model}
%%%%%%%%%%%%%%%%%%%%%%%%%%%%%%%%%%%%%%%%%%%%%%%%%%%%%%%%%%%%%%%%%%

 Here we consider the manifold
 \begin{equation}
   M = \R  \times \R^{n} \label{2.1}
 \end{equation}
 with the metric
 \begin{equation}
  g= -  dt \otimes dt  +
 \sum_{i=1}^{n} e^{2 \beta^i (t)}  dy^i \otimes dy^i,
 \label{2.2}
 \end{equation}
 where $\beta^i (t)$ are smooth functions, $i = 1, \dots, n$.
 
We introduce ``Hubble-like'' variables $h^i = d \beta^i/dt$.
The equations of motion for the action (\ref{act})  read as  follows
   \begin{eqnarray}
        \alpha_1 ( G_{ij} h^i h^j + 2 \Lambda ) 
            -  \alpha_2   G_{ijkl} h^i h^j h^k h^l = 0,  \label{5.1}
         \\
          \left[ 2   \alpha_1  G_{ij} h^j
        -  \frac{4}{3} \alpha_2  G_{ijkl}  h^j h^k h^l \right] \sum_{i=1}^nh^i
        \qquad \nonumber \\
          + \frac{d}{dt} \left[ 2   \alpha_1  G_{ij} h^j
           -  \frac{4}{3} \alpha_2  G_{ijkl}  h^j h^k h^l \right]
           - L    = 0,   \label{5.2}
      \end{eqnarray}
     $i = 1,\ldots, n$, where
      \begin{equation}
       L =  \alpha_1 ( G_{ij} h^i h^j + 2 \Lambda )
                - \frac{1}{3} \alpha_2   G_{ijkl} h^i h^j h^k h^l.
         \label{5.1a}
       \end{equation}
     
     Here
       \begin{eqnarray}
            G_{ij} = \delta_{ij} -1,
              \label{2.10}   \\
            G_{ijkl}  = G_{ij} G_{ik} G_{il} G_{jk} G_{jl} G_{kl}
            \label{2.11}
           \end{eqnarray}
           are respectively the components of two  metrics on
           $\R^{n}$ \cite{Iv-09,Iv-10}. The first one is the well-known ``minisupermetric'' - 2-metric
           of  pseudo-Euclidean signature and the second one is the Finslerian 4-metric.
     
       Due to (\ref{5.1})
      \begin{equation}
       L =   \frac{2}{3}  \alpha_1  (G_{ij} h^i h^j -  4 \Lambda).
        \label{5.1b}
               \end{equation}

  In this paper we deal with the following solutions to
     equations  (\ref{5.1}) and (\ref{5.2})
       \begin{equation}
             h^i (t) = v^i,  \label{5.4v}
       \end{equation}
       with constant $v^i$,   which corresponding to the solutions
      \begin{equation}
      \beta^i = v^i t +       \beta^i_0,  
      \label{5.4a}
      \end{equation}
     where $\beta^i_0$ are constants, $i = 1,\ldots, n$.

      In this case we obtain the metric (\ref{2.2})
      with the exponential dependence of scale  factors
      \begin{equation}
        g= - d t \otimes d t  +
        \sum_{i=1}^{n} B_i e^{2v^i t} dy^i \otimes dy^i,
        \label{5.4m}
        \end{equation}
       where 
        $B_i > 0$ are arbitrary constants.

      For the fixed point  $v = (v^i)$ we have the set of  polynomial equations
          \begin{eqnarray}
         G_{ij} v^i v^j + 2 \Lambda
         - \alpha   G_{ijkl} v^i v^j v^k v^l = 0,  \label{5.5} \\
          \left[ 2   G_{ij} v^j
        -  \frac{4}{3} \alpha  G_{ijkl}  v^j v^k v^l \right] \sum_{i=1}^n v^i
        -  \frac{2}{3}   G_{ij} v^i v^j  +  \frac{8}{3} \Lambda = 0,   \label{5.6}
      \end{eqnarray}
     $i = 1,\ldots, n$, where  $\alpha = \alpha_2/\alpha_1$.
      For $n > 3$ this is a set of forth-order polynomial
      equations.

     For $\Lambda =0$ and $n > 3$ the set of equations (\ref{5.5}) and (\ref{5.6})
      has an isotropic solution $v^1 = ... = v^n = H$,
      only if $\alpha  < 0$  \cite{Iv-09,Iv-10}   $ H = \pm 1 / \sqrt{|\alpha| (n -2)(n -3)}$.
      This solution was generalized in \cite{ChPavTop} to the case
      $\Lambda \neq 0$. 

  It was shown in \cite{Iv-09,Iv-10} that there are no more than
  three different  numbers among  $v^1,...,v^n$ when $\Lambda =0$. This is valid also
  for  $\Lambda \neq 0$.

\section{Examples of Cosmological  Solutions}

 In this section we  consider some
 solutions to the set of equations (\ref{5.5}), (\ref{5.6}) of the following form
  \begin{equation}
 \label{V}
   v =(H, \ldots, H, h, \ldots, h).
 \end{equation} 
 where $H$  the ``Hubble-like'' parameter corresponding  to $m$-dimensional isotropic subspace
with $m > 3$ and $h$ is the ``Hubble-like" parameter   corresponding to $l$-dimensional isotropic subspace, $l>2$.

 These solutions should satisfy  the following conditions:  $H > 0$, $h < 0$.
 The first inequality $H > 0$  is
necessary for  a description of accelerated expansion of
3-dimensional subspace, which may describe our Universe, while the
second inequality $h < 0$ is necessary for contraction of internal space volume.

\subsection{Polynomial equations}

 According to our ansatz (\ref{V}),  we have $m$ dimensions expanding
 with the Hubble parameter $H >0$ and $l$ dimensions contracting  with the ``Hubble-like''  parameter $h <0$.   
 The set of polynomial equations (\ref{5.5}), (\ref{5.6}) reads

\begin{eqnarray} \label{Hh0} 
H^{2}(m-m^{2})+h^{2}(l-l^{2})-2mlHh \nonumber \\
-\alpha(H^{4}m(m-1)(m-2)(m-3)+h^{4}l(l-1)(l-2)(l-3) \nonumber \\
+4H^{3}hm(m-1)(m-2)l+4h^{3}Hl(l-1)(l-2)m \nonumber \\
+6H^{2}h^{2}m(m-1)l(l-1))+ 2 \Lambda = 0, \label{U0}
\end{eqnarray}

\begin{eqnarray} \label{H} 
m(1-m)H^{2}-(1/2)lh^{2}(1+2l)+2lHh((3/4)-m) \nonumber \\
-\alpha(H^{4}m(m-1)(m-2)(m-3)+H^{3}hl(m-1)(m-2)(4m-3) \nonumber \\
+3H^{2}h^{2}l(m-1)(2lm-2l-m) \nonumber \\
+Hh^{3}l(l-1)(4lm-3l-2m)+h^{4}l^{2}(l-1)(l-2)) + 2 \Lambda = 0,
\label{UH}
\end{eqnarray}

\begin{eqnarray}  \label{h} 
l(1-l)h^{2}-(1/2)mH^{2}(1+2m)+2mHh((3/4)-l)\nonumber\\
-\alpha(h^{4}l(l-1)(l-2)(l-3)+h^{3}H m(l-1)(l-2)(4l-3)\nonumber\\
+3h^{2}H^{2} m(l-1)(2lm-2m-l)\nonumber\\+h H^{3}
m(m-1)(4lm-3m-2l)+H^{4}m^{2}(m-1)(m-2)) + 2 \Lambda = 0.
\label{Uh}
\end{eqnarray}

We put   $\alpha = \pm 1$ and denote $\Lambda = \lambda$,  
keeping in mind the general $\alpha$-dependent form of solution
\begin{equation}
\label{HhL}
H(\alpha) = H |\alpha|^{-1/2}, \qquad h(\alpha) = h |\alpha|^{-1/2} 
\qquad  \Lambda = \lambda |\alpha|^{-1}.
\end{equation}

\subsection{Solutions with $\Lambda =0$}

 Let $\Lambda =0$ and $\alpha =  1$.
 It was shown in \cite{IvKob} that, for $m=9$   there exists an infinite series of
 cosmological solutions with $l=3000,3001, ...$, any of which  describes an accelerated
 expansion of the 3-dimensional factor space with sufficiently small  value of
 the variation of the effective gravitational constant $G$ obeying the observational 
 restrictions \cite{Pitjeva}, see also \cite{Ade}. This variation may be arbitrary small
 for a big enough value of $l$. We remind that  the effective
 gravitational constant $G$ is proportional to the inverse volume scale factor
 of the internal space,  see  \cite{BIM,AIKM,IKMN,Mel,IvMel-14} and references therein.

 For $m=11$ and  $l=16$ it was found in  \cite{IvKob} a solution with 
 \begin{equation}
  \label{L02}
   H= \frac{1}{\sqrt{15}}, \qquad  h= -\frac{1}{2 \sqrt{15}},                  
  \end{equation}
  which describe a zero variation of effective cosmological constant $G$.
 
 Another solution of such type which was found in  \cite{IvKob} appears for  $ m=15$ and $l=6$ with
 \begin{equation}
 \label{L03}
  H=\frac{1}{6}, \qquad  h=-\frac{1}{3}.                  
 \end{equation}

\subsection{Solutions with $\Lambda \neq 0$}

 Here we present several new cosmological solutions for $\Lambda \neq 0$, $\alpha =  1$, $m =3$ and $l=4$.
 
 The first solution takes place for $\lambda = 3/16$:
  \begin{equation}
 \label{l1}
 H = \frac{1}{4} \sqrt{2}, \qquad   h = - \frac{1}{4} \sqrt{2}.
 \end{equation}
 
  The second one is valid for $\lambda = 13/48$:
   \begin{equation}
   \label{l2}
   H = \frac{1}{4} \sqrt{6}, \qquad   h = - \frac{1}{12} \sqrt{6}.
   \end{equation}
 
  The third one 
    \begin{equation}
    \label{l3}
      H = \frac{2}{29} \sqrt{29}, \qquad   h = - \frac{3}{58} \sqrt{29}.
      \end{equation}
   corresponds to $\lambda = 21/116$.
 
 Any of these solutions describe accelerated expansion of 3-dimensional factor
space and contraction of internal space. All of these solutions take place for
fixed positive values of $\lambda$ (see (\ref{HhL})). 
There exist also examples of solutions with negative cosmological constant $\lambda = - 21/80$:
 \begin{eqnarray}
    \label{l4}
      H_{\pm} = \frac{1}{60320} ( 248 \pm 32 \sqrt{30}) \sqrt{68150 \mp 9280 \sqrt{30}}, \\
      h_{\pm} = - \frac{1}{580}  \sqrt{68150 \mp 9280 \sqrt{30}}.
     \label{l5}
      \end{eqnarray}
 These solutions obey inequalities  $H_{\pm} > 0$ and  $h_{\pm} < 0$.

 \section{Conclusions}

 We have considered the  $D$-dimensional  Einstein-Gauss-Bonnet (EGB) model with
 with $\Lambda$-term.
 By using the  ansatz with diagonal  cosmological type metrics,
 we have found new solutions with exponential dependence of scale factors
 with respect to synchronous time variable $t$ in dimension $D = 1 + 3 + 4$.
Any of these   solutions   describes   an exponential expansion of ``our'' 3-dimensional factor-space with
 the Hubble parameter $H > 0$ and exponential contraction of $4d$ internal space.
 An open question arising here is to find solutions with $\Lambda \neq 0$ which  
 obey  the observational constraints on the temporal variation of the effective gravitational constant $G$.
 This question will be addressed in a separate publication.

\end{document}